\documentclass[11pt]{article}
\usepackage{geometry}
\usepackage[parfill]{parskip}
\usepackage{graphicx}
\usepackage{amssymb}
\begin{document}

\textbf{\Large{Spatial multimode entanglement within one laser beam}}
\vspace{.25in}

J. Janousek$^{1\ast}$, K. Wagner$^{1}$, J-F. Morizur$^{1,2}$,
N. Treps$^{2}$, P. K. Lam$^{1}$,\\ C. C. Harb$^{3}$, H-A.
Bachor$^{1}$

\normalsize{$^{1}$ARC Centre of Excellence for Quantum-Atom Optics}\\
\normalsize{The Australian National University, Canberra ACT 0200, AUSTRALIA}\\
\normalsize{$^{2}$Laboratoire Kastler Brossel, Paris Cedex 5, FRANCE}\\
\normalsize{$^{3}$Australian Defence Force Academy, Canberra, AUSTRALIA}

\normalsize{$^\ast$To whom correspondence should be addressed:     jiri.janousek@anu.edu.au}

\vspace{.25in}

\textbf{Optical entanglement is a key requirement for many quantum communication protocols. Conventionally entanglement is formed between two distinct beams, with the quantum correlations being measured at separate locations. We show entanglement between the modes within one beam. Our technique is particularly elegant and a major advance towards practical systems with minimum complexity.  We demonstrate three major experimental achievements: (i) only one source is required to produce squeezed light in two orthogonal spatial modes, (ii) the entanglement is formed through lenses and beam rotation, without the need of a beam splitter and (iii) the quantum correlations are measured directly and simultaneously using a multi pixel, quadrant detector. }

Optical entanglement between two beams has been used to study the
fundamental quantum properties of light \cite{reid:09} and for the
demonstration of quantum communication protocols.  The detection
of continuous variables, the amplitude and phase quadrature, has
been used to show dense coding \cite{Lee:02}, teleportation
\cite{Bowen:03,Yonezawa:04}, quantum secret sharing
\cite{Lance:05}  and entanglement distillation
\cite{Dong:08,Hage:08}. We have extended this to the spatial
domain demonstrating EPR correlations between position and
momentum of the photons in two laser beams \cite{Wagner:08}.

Multimode entanglement allows more complex processes and leads to
more advanced techniques. Tripartite GHz correlations and more
recently cluster states, combining four individual squeezed modes,
have been demonstrated with impressive reliability
\cite{Yukawa:08,Su:07}. However, using separate beams to build the
quantum state requires combining complex resources, in particular
many squeezers, beam splitters, phase shifters and a set of
separate homodyne detectors. This technology is difficult to
simplify as it is very sensitive to losses and any mode-mismatch.

An alternative approach is to consider multiple modes within a
single beam.  There have been proposals to use correlated
frequency sidebands generated in one source \cite{Menicucci:07},
to correlate frequency sidebands  \cite{Huntington:05} or to use
temporal modes that describes different pulse shapes
\cite{Valcarcel:06}.  Spatial modes, on the basis of
Hermite-Gaussian (H-G) modes TEM$_{nm}$, can be generated
efficiently, superposed with low losses  \cite{Treps:02,Lassen:07}
and many modes can be measured simultaneously using a multi pixel
homodyne detection \cite{Beck:00,Delaubert:07}. Shaping the local
oscillator using a spatial light modulator and varying the gains
on the detectors changes the measurement basis. This creates a
family of entangled measurements.   As an in principle
demonstration we report here the entanglement of two spatial
modes, TEM$_{01}$ and TEM$_{10}$, within one beam.




\begin{figure}[t]
   \begin{center}
   \includegraphics[width=10.5cm]{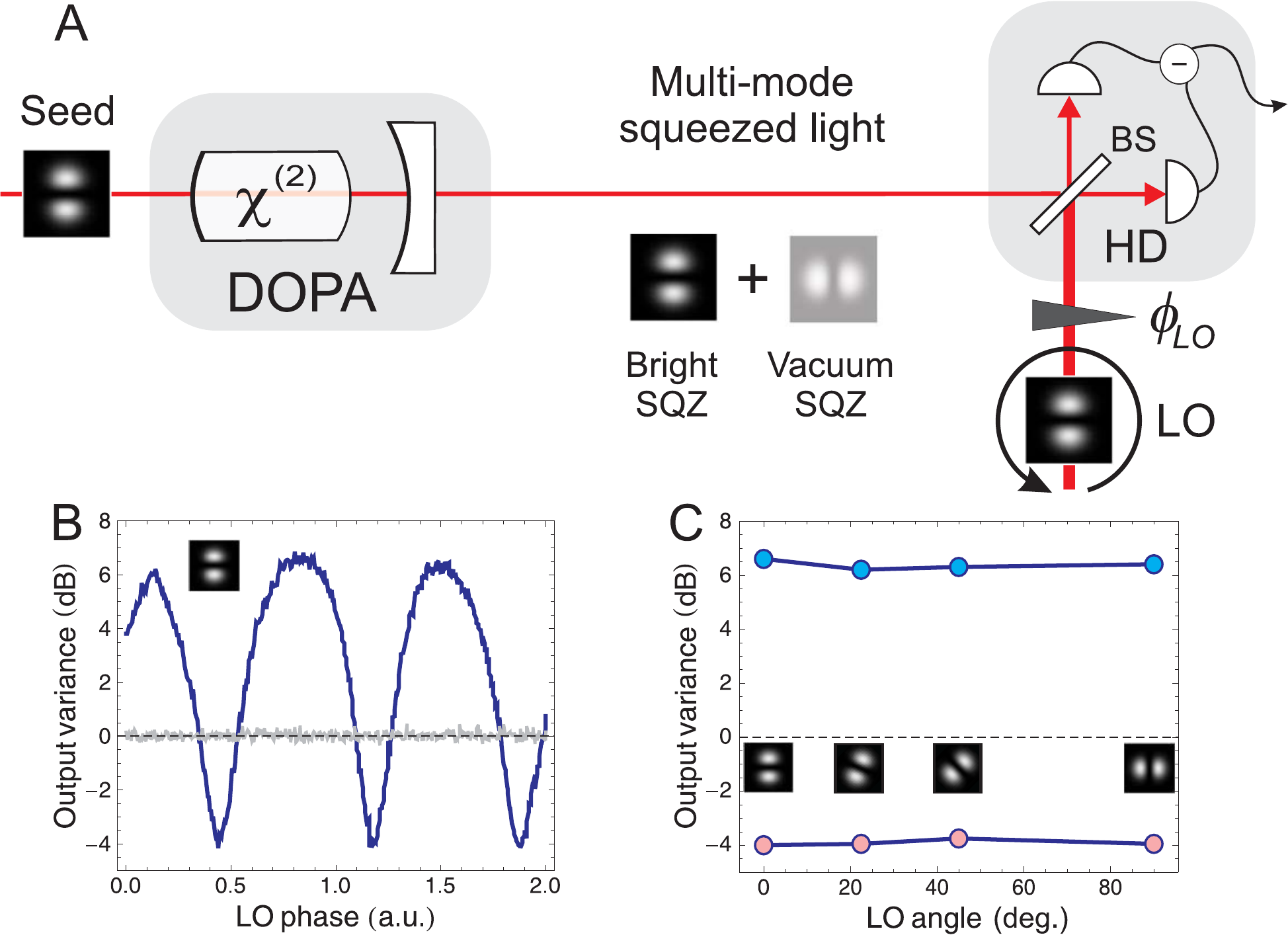}
   \end{center}
    \caption{\textbf{A}: Schematic of the squeezing experimental setup.
    DOPA: degenerate optical parametric amplifier; LO: local oscillator;
    HD: homodyne detection; BS: 50/50 beamsplitter. \textbf{B}: Measurement
    of squeezed field in TEM$_{10}$ mode and scanning the phase of LO.
    \textbf{C}: Measurements of squeezing/antisqueezing when the TEM$_{10}$
    LO beam is spatially rotated using a Dove prism along the $x$-axis.}
    \label{opa_setup}
\end{figure}

The crucial resource required in creating entangled beams is the squeezed light source.  Here we use the process of optical parametric amplification (OPA).  There are several OPA designs, the most common of which are the linear and bow-tie geometries, and we use the former design in order to benefit from the natural degeneracy of optical resonators with cylindrical symmetry.

Squeezed light in two orthogonal modes is produced using a
degenerate OPA operating below threshold, see Fig.
\ref{opa_setup}A. The OPA is pumped with 532 nm light from a
frequency-doubled diode-pumped Nd:YAG laser operating at 1064 nm.
The OPA crystal has dimensions $2 \times 2.5 \times 6.5$ mm$^{3}$
and is made from bulk LiNbO$_{3}$ which is 7\% doped with MgO and
phase-matched at 61$^{o}\mathrm{C}$. The OPA cavity is linear and
is formed by the rear surface of the crystal (radius of curvature
= 8 mm, high reflector at 532 nm, R=99.9\% at 1064 nm) and an
external mirror (radius of curvature = 75 mm, R=13\% at 532 nm and
R=96\% at 1064 nm).  The front surface of the crystal has a radius
of curvature of 8~mm and is anti-reflection coated at both 1064~nm
and 532~nm. The optical path length of the OPA cavity is
approximately 38~mm.

The OPA is seeded with a weak TEM$_{10}$ field incident on the
high-reflecting side of the OPA crystal. The system is carefully
aligned such that the two orthogonal modes, i.e. TEM$_{10}$ mode
and TEM$_{01}$ mode, resonate simultaneously. This degeneracy is
achieved by changing the temperature of the laser crystal. The
system is operated as a de-amplifier with gains of 0.4 using
180~mW of pump power in the TEM$_{00}$ mode. This pump mode is not
an optimum mode for an OPA operating in TEM$_{10}$ mode, but still
provides sufficient nonlinear gain. As a result multimode squeezed
light, a low intensity amplitude squeezed field in TEM$_{10}$ mode
and vacuum squeezed field in TEM$_{01}$ mode, is produced. For an
ideal situation these two fields should be exactly in-phase. The
squeezed  light  is analyzed using homodyne detection (HD) with
the local oscillator (LO) in the TEM$_{10}$ mode. We use a Dove
prism for the spatial rotation of the LO beam in order to analyze
output of the degenerate OPA in any direction, i.e. going from
TEM$_{10}$ to TEM$_{01}$ mode and in-between. For TEM$_{10}$
operation we observed typically  -4~dB of squeezing and +6.5~dB of
anti-squeezing, see Fig.~\ref{opa_setup}B. More interestingly we
observed states of approximately the same squeezing and
antisqueezing when the LO TEM$_{10}$ mode is rotated in respect to
the $x$-axis, see Fig.~\ref{opa_setup}C. This clearly demonstrates
multi-mode squeezing generation using a single linear degenerate
OPA.

\begin{figure}[htbp]
   \begin{center}
   \includegraphics[width=11.5cm]{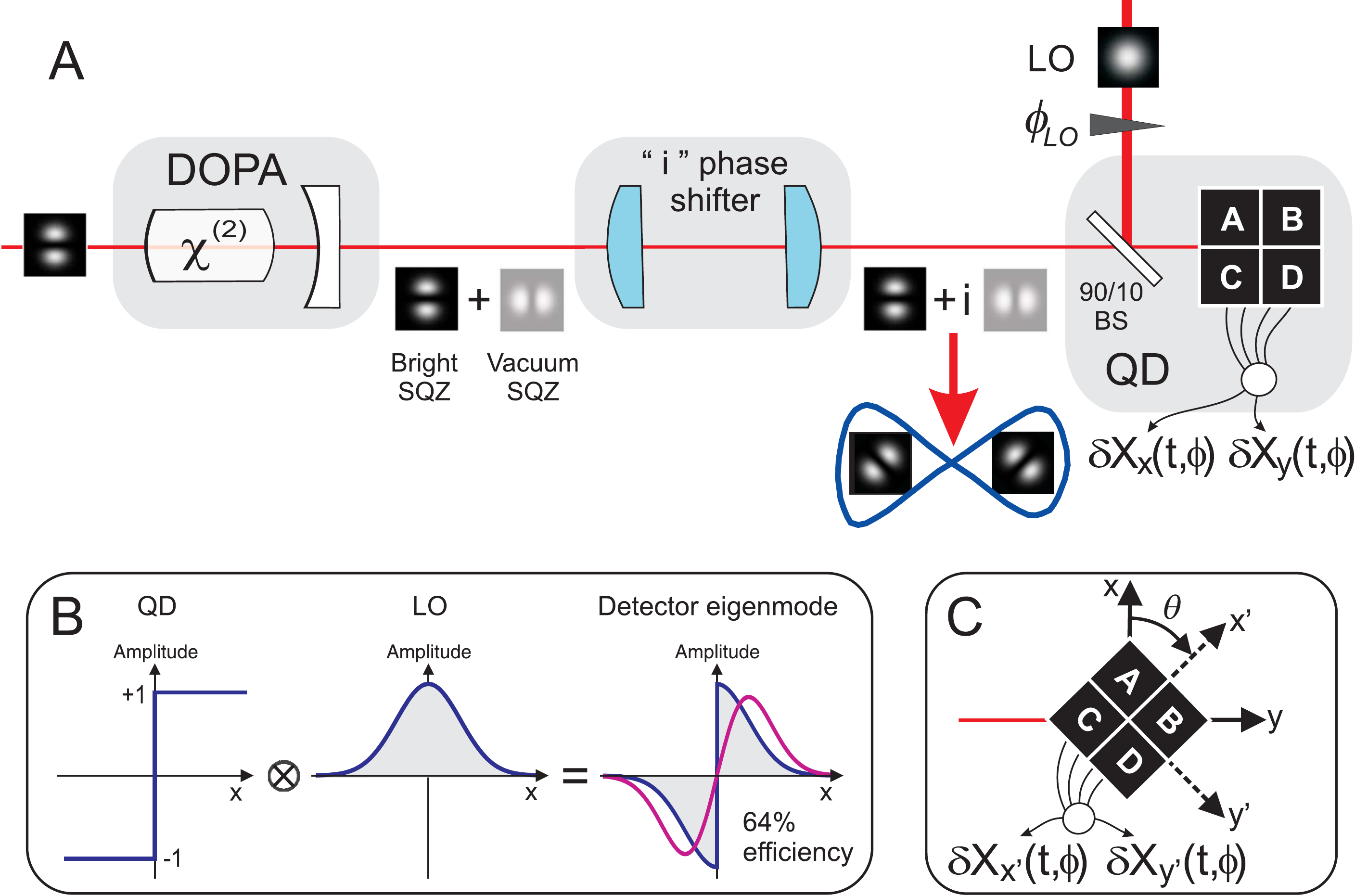}
   \end{center}
    \caption{\textbf{A}: Multi-mode entanglement experimental setup.  $\delta X_{x}\left(t,\phi\right)$ is equivalent to $\delta X_{\left(A+B\right)-\left(C+D\right)}\left(t,\phi\right)$, and $\delta X_{y}\left(t,\phi\right)$ is given by $\delta X_{\left(A+C\right)-\left(B+D\right)}\left(t,\phi\right)$.
    DOPA: degenerate optical parametric amplifier; LO: local oscillator;
    HD: homodyne detection; QD: quadrant detector. \textbf{B}: Principle of
    split-detection technique. The eigenmode of a split-detector is a flipped
    mode, resulting in 64\% detection efficiency in TEM$_{10}$ basis.
    \textbf{C}: Spatial 50/50 beamsplitter is introduced just by a measurement
    in a 45$^{\circ}$ rotated basis.}
    \label{multi-ent_setup}
\end{figure}

\begin{figure}[t]
   \begin{center}
   \includegraphics[width=12cm]{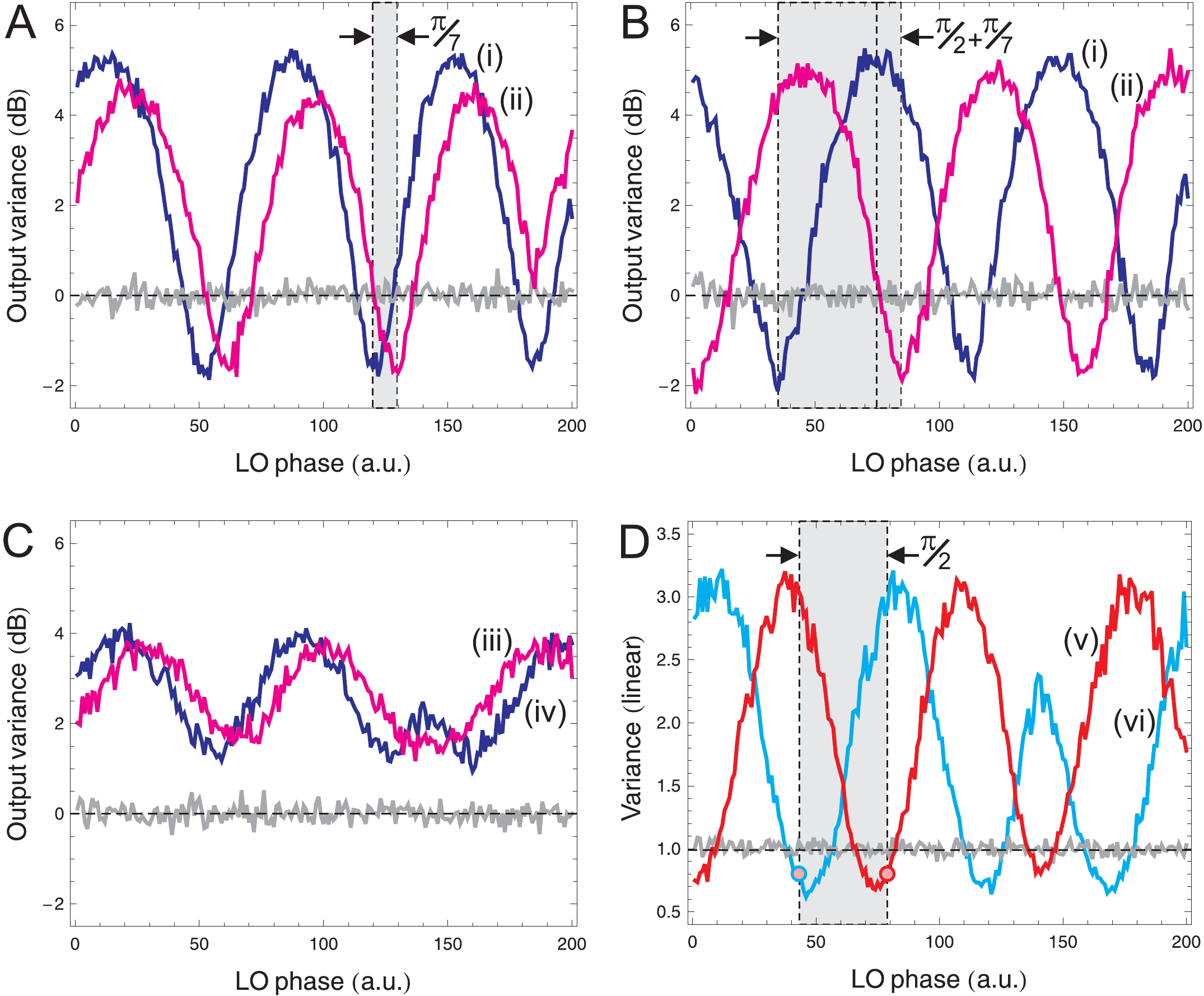}
   \end{center}
    \caption{ \textbf{A}: Output variance of the degenerate OPA for TEM$_{10}$ field, with variance $V_{x}\left(\phi\right)$
    (i), and TEM$_{01}$ field, with variance $V_{y}\left(\phi\right)$ (ii), using a QD and scanning LO phase; \textbf{B}: The same setup as  \textbf{A}, but with the cylindrical-lens system included. \textbf{C}:  Output variances $V_{x'}\left(\phi\right)$ (iii) and $V_{y'}\left(\phi\right)$ (iv) for the 45$^{\circ}$ rotated fields using a QD and scanning the LO phase. \textbf{D}: Measurement of the variance of  sum $V_{x'+y'}\left(\phi\right)$ (v) and
    difference $V_{x'-y'}\left(\phi+\frac{\pi}{2}\right)$ (vi) photocurrents for the 45$^{\circ}$ rotated fields.}
    \label{measurements}
\end{figure}

In order to prove that the two orthogonal states produced by the
degenerate OPA are  independent we used a quadrant detector (QD)
with one LO field in the TEM$_{00}$ mode, see
Fig.~\ref{multi-ent_setup}A. The eigen-mode of such a detection
system is a flipped mode, see Fig.~\ref{multi-ent_setup}B, giving
us at the best only 64\% efficiency for TEM$_{10}$ detection.
However, by combining the outputs of the QD with electronic
splitters and using a fast data acquisition system, we are able to
measure properties of states in the TEM$_{10}$ mode and TEM$_{01}$
mode simultaneously, using just a single detector and one local
oscillator. The temporal fluctuations are directly recorded, then
the data is post-processed. It is filtered for a certain frequency
(here 4.8 MHz, with a width of 100kHz), and the correlations and
variances are then directly determined from the time varying data.
Such measurements are shown in Fig.~\ref{measurements}A. After
detection on the quadrant detector -1.7~dB of squeezing is
measured in the two modes, which is sufficient for a clear
demonstration of entanglement generation. An interesting feature
and limiting factor is the small phase shift, about  $\pi/7$,
between the two fields, which might have an origin in a small
misalignment of the OPA cavity.

There is a well-established set of requirements for entangling two
optical beams, and we meet all of these in our unusual setup. A
$\pi/2$ (or $i$) phase shift is first required between the two
beams, which for standard entanglement is simply a matter of
delaying one of the beams with respect to the other.  The beams
then need to be mixed together, which is generally achieved with a
50/50 beamsplitter. Finally, we need to observe a pair of
conjugate observables, which requires a phase-sensitive detector
in order to measure quadrature entanglement.  This is usually
achieved with one homodyne detector on each of the entangled
beams. To induce the $\pi/2$ phase shift we used an elegant
optical method employing the Gouy phase shift in higher-order
modes \cite{Siegman:86}. The output of the degenerate OPA was
mode-matched into a symmetric two cylindrical-lens system (focal
lengths $f$=250~mm, with lens separation of $\sqrt{2} f$), which
results in the required phase shift, as shown in the comparison of
the squeezing results in Fig.~\ref{measurements}A and
Fig.~\ref{measurements}B.

The last missing piece in the experiment is the equivalent to a
50/50 beam-splitter, in order to mix the TEM$_{10}$ and TEM$_{01}$
modes. Any H-G mode can be expressed as a superposition of two
orthogonal modes of the same order as the original field. This is
analogous to the superposition polarization modes in a 2
dimensional basis. A TEM$_{10}$ mode rotated by 45$^{\circ}$
relative to the $x$-axis can be expressed as
$\frac{1}{\sqrt{2}}$TEM$_{10}\pm\frac{1}{\sqrt{2}}$TEM$_{01}$.
This means that our `beam-splitter' can be realized by detecting a
basis which is 45$^{\circ}$ rotated  relative to the axis of the
cylindrical-lens system, see Fig.~\ref{multi-ent_setup}C. As
expected from quantum theory, measurements of the arbitrary
quadratures of entangled fields show noisy states, see
Fig.~\ref{measurements}C. However, taking the sum and difference
of the signals of the two orthogonal fields shows a clear
signature of entanglement, as seen in Fig.~\ref{measurements}D.

With these data we can calculate the Inseparability criterion
$\mathcal{I}$, which is a direct measure of entanglement. We
evaluate the equation
$\mathcal{I}=\sqrt{V_{x'+y'}\left({\phi_{o}}\right)
V_{x'-y'}\left({\phi_{o}+\frac{\pi}{2}}\right)}$, where $\phi_{0}$
is chosen such that $\mathcal{I}$ is minimized.  This gives us a
value of $\mathcal{I}=0.81$, after corrections are made for
electronic noise, and thereby demonstrate entanglement between two
orthogonal H-G modes within one optical beam. This concept can be
used to produce entanglement between any two orthogonal H-G modes
of the same order.

In conclusion, we have demonstrated an elegant technique to create and measure entanglement between two orthogonal spatial modes in a single beam of light.  We have shown several simplifications on traditional entanglement schemes, including generating two squeezed modes from a single OPA, using imaging components to mix the modes with the correct phase and detecting the two modes simultaneously with one quadrant detector.  This technique can be expanded to several higher order modes more TEM$_{nm}$. We can synthesize a beams from several sources of squeezed, mix the modes using imaging techniques and detect the orthogonal modes using multi-pixel detectors and one local oscillator.    Using this infinite basis within a single beam and the possible manipulation of the modes makes such an optical communication system of interest for quantum protocols.


\vspace{1cm} \textbf{Acknowledgements}

This work was funded by the Centre of Excellence program of the
Australian Research Council and was supported CNRS and the Ecole
Normale Superieur, Paris. We would like to thank Anais Dreau for
her contributions to design of the data aquisition system.


\end{document}